\RequirePackage{fix-cm} % Fix LaTeX2e bugs.
\documentclass[a4paper, twoside, 12pt, dvips]{amsart}

\usepackage{fixltx2e}     % Fix LaTeX2e bugs.

\usepackage{amssymb}
\usepackage{amsmath}

\usepackage{a4wide}

\usepackage{indentfirst}
\usepackage{graphicx}
\usepackage{psfrag}

\usepackage[usenames,dvipsnames]{pstricks}
\usepackage{epsfig}
\usepackage{pst-grad} % For gradients
\usepackage{pst-plot} % For axes

\usepackage{subfigure}

\usepackage{longtable}

\usepackage[english]{babel}
\usepackage[latin1]{inputenc}

\usepackage{color}
\definecolor{oneblue}{rgb}{0,0.0,0.75}
\usepackage[colorlinks,
            urlcolor=oneblue,
            linkcolor=oneblue,
            citecolor=oneblue,
            bookmarksopen=false,
            pagebackref]{hyperref}

\numberwithin{equation}{section}

\newtheorem{rem}{Remark}

\newcommand{\dt}{\partial_t}

\newcommand{\abs}[1]{\left|#1\right|}

\def\g{\vec{g}}
\def\u{\vec{u}}
\def\n{\vec{n}}
\def\v{\vec{v}}
\def\x{\vec{x}}
\def\D{\mathbb{D}}
\def\A{\mathbb{A}}
\def\nub{\bar{\nu}}

\def\div{\nabla\cdot}
\def\ttau{\vec{\tau}}
\def\grad{\nabla}

\def\Sc{\mathrm{Sc}}
\def\Ma{\mathrm{Ma}}
\def\At{\mathrm{At}}
\def\Re{\mathrm{Re}}
\def\Fr{\mathrm{Fr}}

\def\OpenFOAM{\textsf{OpenFOAM }}

\begin{document}

\title{Mathematical modeling of powder-snow avalanche flows}

\author[D. Dutykh]{Denys Dutykh$^*$}
\address{LAMA, UMR 5127 CNRS, Universit\'e de Savoie, Campus Scientifique,
73376 Le Bourget-du-Lac Cedex, France}
\email{Denys.Dutykh@univ-savoie.fr}
\urladdr{http://www.lama.univ-savoie.fr/~dutykh/}
\thanks{$^*$ Corresponding author}

\author[C. Acary-Robert]{C\'eline Acary-Robert}
\address{LAMA, UMR 5127 CNRS, Universit\'e de Savoie, Campus Scientifique,
73376 Le Bourget-du-Lac Cedex, France}
\email{Celine.Acary-Robert@univ-savoie.fr}
\urladdr{http://www.lama.univ-savoie.fr/~acary-robert/}

\author[D. Bresch]{Didier Bresch}
\address[D. Bresch]{LAMA, UMR 5127 CNRS, Universit\'e de Savoie, 73376 Le Bourget-du-Lac Cedex, France}
\email{Didier.Bresch@univ-savoie.fr}

\begin{abstract}
Powder-snow avalanches are violent natural disasters which represent a major risk for infrastructures and populations in mountain regions. In this study we present a novel model for the simulation of avalanches in the aerosol regime. The second scope of this study is to get more insight into the interaction process between an avalanche and a rigid obstacle. An incompressible model of two miscible fluids can be successfully employed in this type of problems. We allow for mass diffusion between two phases according to the Fick's law. The governing equations are discretized with a contemporary fully implicit finite volume scheme. The solver is able to deal with arbitrary density ratios. Several numerical results are presented. Volume fraction, velocity and pressure fields are presented and discussed. Finally we point out how this methodology can be used for practical problems.
\end{abstract}


\maketitle

\section{Introduction}

\begin{figure}
  \centering
    \subfigure{\includegraphics[width=0.49\textwidth]{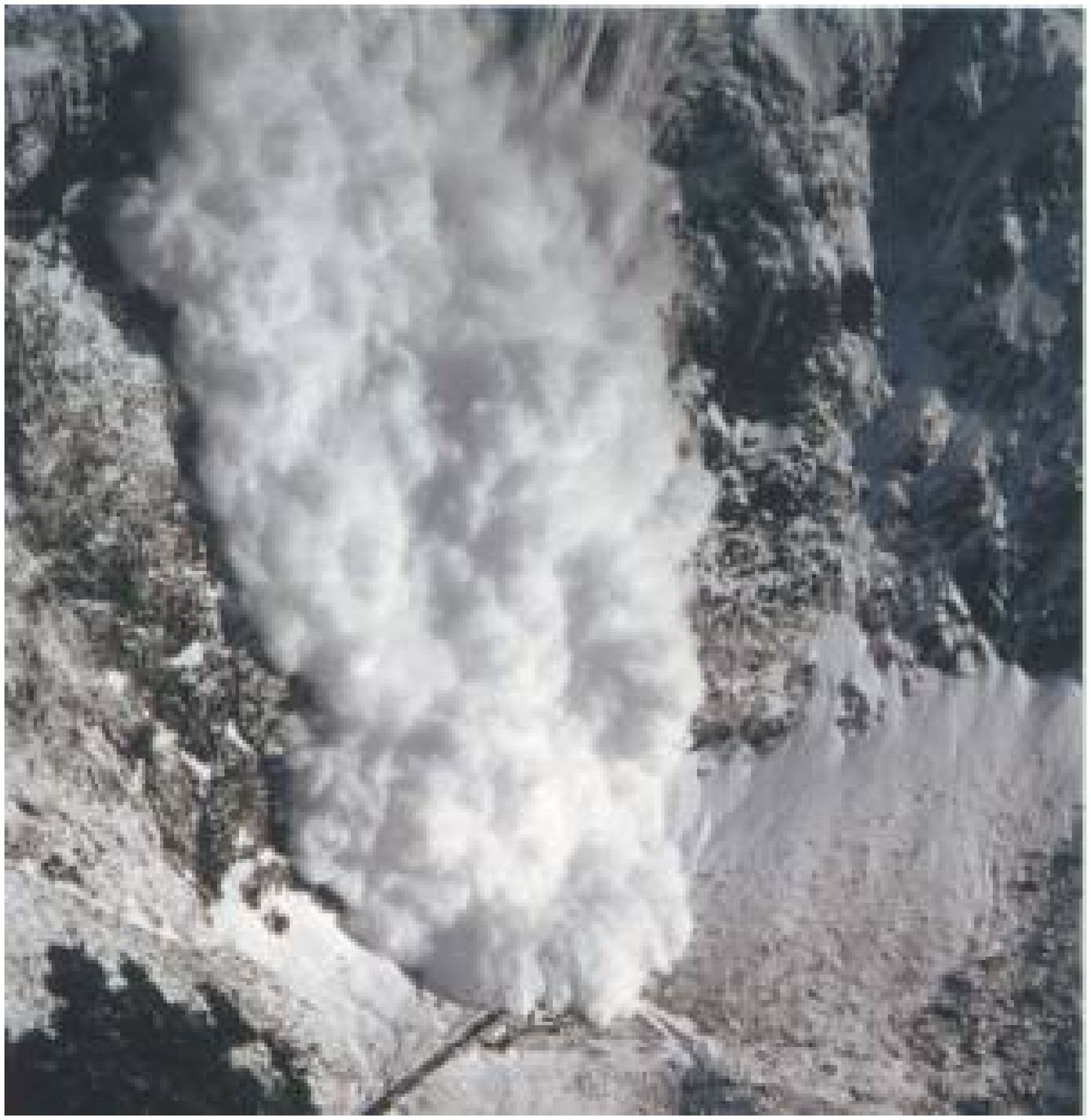}}
    \subfigure{\includegraphics[width=0.4\textwidth]{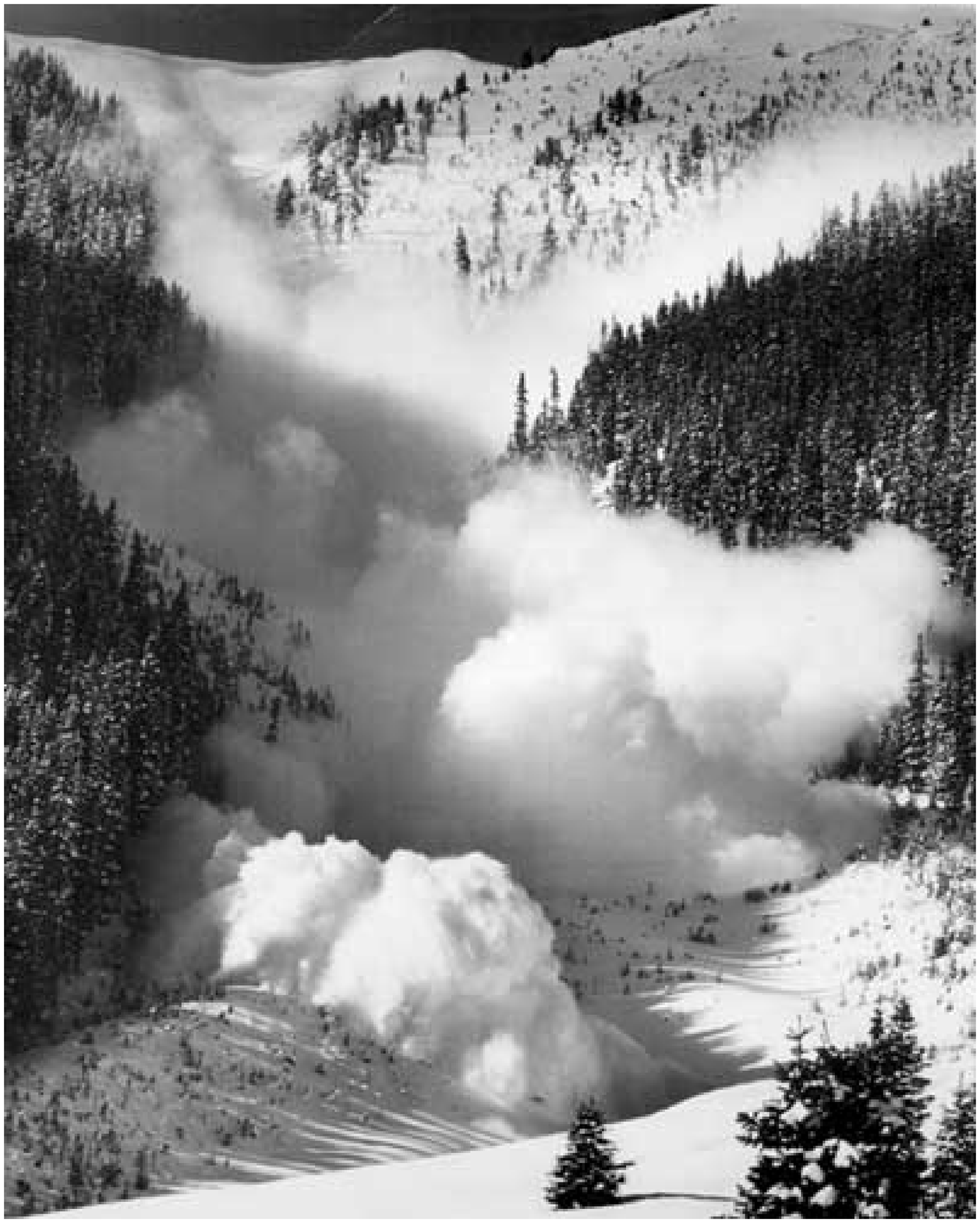}}
  \caption{Two illustrations of powder-snow avalanche flows.}
  \label{fig:aval}
\end{figure}

Snow avalanches are commonly defined as abrupt and rapid gravity-driven flows of snow down a mountainside, often mixed with air, water and sometimes debris (see Figure \ref{fig:aval}). Avalanches are physical phenomena of great interest, mainly because they represent a big risk for those who live or visit areas where this natural disaster can occur. During last several decades, the risk has increased due to important recreational and construction activities in high altitude areas. We remember recent events at Val d'Is\`ere in 1970 and in Northern Alps in 1999 \cite{Ancey2001}.

The avalanches arise from an instability in a pile of granular material like sand or snow \cite{Turnbull2008}. The destabilization phase of an avalanche life is still a challenging problem. There are many factors which influence the release process. One can recall snowpack structure, liquid contain, shape and curvature of starting zone and many others \cite{Ancey2001}. In this study we focus especially on sliding and stopping phases.

The serious research work on this natural phenomenon was preceded by the creation of scientific nivology at the end of the XIXth century. Among the pioneers we can mention Johann Coaz (swiss engineer) \cite{Coaz1881} and P.~Mougin (french forest engineer, author of the first avalanche model using an analogy with a sliding block) \cite{Mougin1922, Mougin1931}. The work of P.~Mougin was ignored until the 1950s when A. Voellmy developed a similar model \cite{Voellmy1955}. Its somehow improved versions are still used by engineers nowadays.

We would like to point out here important contributions of the Soviet school by S.S. Grigorian, M.E. Eglit, A.G. Kulikovskiy, Y.L. Yakimov and many others \cite{Grigorian1967, Kulikovskiy1973, Bakhvalov1973, Bakhvalov1975, Kulikovskiy1977, Eglit1983, Eglit1991, Bozhinskiy1998, Eglit1998, Blagovechshenskiy2002}. They were at the origin of all modern avalanche models used nowadays in the engineering practice and, sometimes, in scientific research. Their works were mainly devoted to the derivation and comprehension of mathematical models while occidental scientists essentially looked for quantitative results.

Conventionally we can divide all avalanches in two idealized types of motion: flowing and powder-snow avalanches. A flowing avalanche is characterized by a high-density core ranging from $100$ to $500$ $kg/m^3$ and consists of various types of snow: pasty, granular, slush, etc. The flow depth is typically about a few meters which is much smaller than the horizontal extent. This argument is often used to justify numerous depth-integrated models of the Savage-Hutter type\footnote{In hydrodynamics and hydraulics this type of modeling is also known as shallow water or Saint-Venant equations \cite{SV1871}.} \cite{Savage1989, Savage1991}. These avalanches can cause extensive damage because of the important snow masses involved in the flow in spite of their low speed.

\begin{figure}
	\centering
		\includegraphics[width=0.59\textwidth]{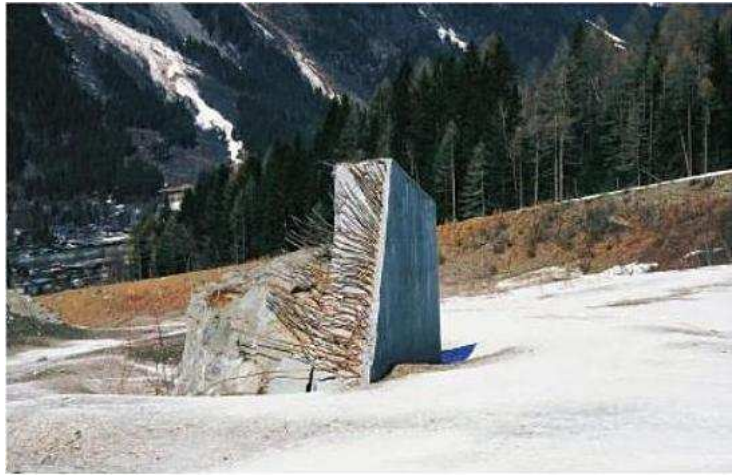}
	\caption{Protecting wall in armed concrete at Taconnaz (Haute-Savoie, France) destroyed by the powder-snow avalanche of the 11 February 1999. The height is $7$ m and the thickness is $1.5$ m (Photo by C. Ancey).}
	\label{fig:dentbeton}
\end{figure}

On the other hand, powder-snow avalanches are large-scale turbidity currents descending slopes at high velocities \cite{Rastello2004, Ancey2006a, Ancey2007}. They seriously differ from flowing avalanches. These clouds can reach $100$ m in height and very high front velocities of the order of $100$ m/s. They grow continously and the average density is fairly low (from $4$ to $25$ $kg/m^3$). These spectacular avalanches (see Figure \ref{fig:aval}) occur only under certain conditions (after abundant fresh snowfalls, cold, dry and weakly cohesive snow on strong slopes) and they produce a devastating pressure wave which breaks the trees, buildings, tears off the roofs, etc. During the propagation stage, they are able to cross the valleys and even to climb up on the opposite slope. Hence, measurements by intrusive probes are almost impossible. Avalanches in aerosol are not very frequent events in Alps but in the same time we cannot say they are very seldom. In the technical literature there is an opinion that an avalanche in aerosol is less destructive than a flowing one since the transported mass is much smaller. Nevertheless, recent events of the winter 1999 in Switzerland, Austria and France revealed the important destructive potential of the powder-snow avalanches (see Figure \ref{fig:dentbeton}).

Recently several systematic measurements campaigns \textit{in situ} were conducted in Norway, Switzerland and Japan \cite{McClung1984, Norem1990, Nishimura1993, Nishimura1995, Dufour2000}. Researchers shed some light on the internal structure of big avalanches. More precisely, they show that there exists a dense part of the avalanche which remains permanently in contact with the bed. This dense core is covered by the aerosol suspension of snow particles in the air. From these results it follows that mentioned above two types of avalanches may coexist in nature and proposed above classification is rather conventional. Perhaps, future studies will perform a coupling between the dense core and powder-snow envelope in the spirit of SAMOS code \cite{Sampl2004}.

Let us review various existing approaches to the mathematical modeling of snow avalanches. Generally, we have two big classes of mathematical models: probabilistic and deterministic. In the present article we deal with a deterministic model and we refer to the works of K. Lied and D.M. McClung \cite{Lied1980, McClung1987, McClung2000, McClung2001} for more information on statistical approaches to avalanche modeling. Deterministic models can be further divided into continuous and discrete ones depending whether the material under consideration can be approximated as a continuum medium or not. For the review and some recent results on dense granular flows we refer to following works \cite{Nishimura1998, McElwaine2001, Rajchenbach2002, Rajchenbach2002a, Rajchenbach2005} and the references therein. Some promising results were obtained with discrete models based on cellular automata \cite{DiGregorio1999, Avolio2000, D'Ambrosio2006}.

The first contemporary avalanche models appeared in 1970 by soviet scientists Kulikovskiy and Sveshnikova \cite{Kulikovskiy1977, Bozhinskiy1998}. Later, their idea was exploited by Beghin \cite{Beghin1979, Beghin1983, Beghin1991} and others \cite{Hopfinger1977, Fukushima1990, Akiyama1999, Ancey2004, Rastello2004}. We call this type of modeling $0$D-models since the avalanche is assimilated to semi-elliptic cloud with variable in time volume $V(t)$, momentum $(\rho U)(t)$, etc. All quantities of interest are assimilated to the center of mass and their dynamics is governed by conservation laws expressed as Ordinary Differential Equations (ODE).

On the next complexity level we have various depth-integrated models. The governing equations are of Shallow Water (or Saint-Venant \cite{SV1871}) type. In general, they are derived by depth averaging process or some asymptotic expansion procedure from complete set of equations. Thus, a physical $3$D (or $2$D) problem results in a $2$D model ($1$D correspondingly). From computational point of view these models are very affordable even for desktop computers. On the other hand, they provide us very approximative flow structure, especially in the vertical direction.

In France, G. Brugnot and R. Pochat were among the pioneers \cite{Brugnot1981} while in Soviet Union this direction was explored in the beginning of 1970 by N.S. Bakhvalov et M.E. Eglit \cite{Bakhvalov1973, Bakhvalov1975}. Currently, each country concerned with the avalanche hazard, has its own code based on this type of equations. Probably the most representative model is that developed by S. Savage and K. Hutter \cite{Savage1989}. Nowadays this set of equations is generally referred as the Savage-Hutter model. We refer to \cite{Hopfinger1983, Hutter1996} as general good reviews of existing theoretical models and laboratory experiments.

This approach was further developed by incorporating more complex rheologies and friction laws \cite{Gray1998, Hutter1991, Hutter1993, Mangeney-Castelnau2003, Hutter2005, Fernandez-Nieto2007}. To conclude on this part of our review, we have to say that this modeling is more relevant to the flowing avalanche regime which is characterized by the small ratio of the flow depth $h$ to the horizontal extent $\ell$ (i.e. $\frac{h}{\ell} \ll 1$).

Finally, we come to the so-called \textit{two-fluid} (or \textit{two-phase}) models. In this paradigm both phases are resolved and, a priori, no assumption is made on the shallowness of the flow under consideration. Another advantage consists in fact that efforts exerted by the ambient air on the sliding mass are naturally taken into account. From computational point of view, these models are the most expensive \cite{Naaim1998, Etienne2004, Etienne2004a, Etienne2005}. In the same time, they offer quite complete information on the flow structure.

As it follows from the title, in this study we are mainly concerned with powder-snow avalanches. We would like to underline that our modeling paradigm allows for taking into account the dense core at the first approximation order. The density is completely determined by the snow volume fraction distribution. This parameter can be used to introduce a stratification in the initial condition, for example. Otherwise, it will happen automatically due to mixing and sedimentation processes in the flow, provided we follow its evolution for sufficiently long time.

We retained a simple incompressible two-fluid model which is described in detail in Section \ref{sec:model}. Two phases are allowed to interpenetrate, forming a mixing zone in the vicinity of the interface. We make no Boussinesq-type hypothesis \cite{Ancey2004} about the density ratio. Moreover, our solver is robust and is able to deal with high density ratios (we tested up to $3000$). In nature, the flow under consideration is obviously highly turbulent but at the present stage we do not incorporate any explicit turbulence modeling beyond resolved scales. In the present study we focus mainly on the physically sound description of incompressible highly inhomogeneous two-phase flows.

Good understanding of these natural phenomena can improve the risk assessment of such natural hazards. Large scale experiments are not feasible\footnote{However, we would like to notice that there are two experimental sites: one in Switzerland (Sion Valley) and another one in France (Col d'Ornon). Unfortunately, field measurements provide very limited information on the flow structure \cite{Dufour2001}, even if some a posteriori analysis may be beneficial \cite{Turnbull2007a}.}. Field measurements during the event are very hazardous and the events are rare. Laboratory experiments on powder-snow avalanches are essentially limited to Boussinesq clouds\footnote{The Boussinesq regime corresponds to the situation when $\frac{\rho^+ - \rho^-}{\rho^+} \ll 1$ where $\rho^\pm$ are densities of the heavy and light fluids respectively. This asymptotics allows to introduce the so-called Boussinesq approximation.} \cite{Keller1995, Naaim-Bouvet2002, Primus2003, Naaim-Bouvet2003, Primus2004} while it is not the case in nature. Thus, we do not really know how these results apply to real-world events. Fortunately, some progress has been made recently to remedy this situation \cite{Turnbull2008}.

Numerical simulations of avalanches provide useful information on the dynamics of these flows. Computer experiments may become in term the main tool in testing various situations. Experiments \textit{in silico} should be complementary to those \textit{in situ} or in laboratory. Direct Numerical Simulations (DNS) provide complete information about all flow quantities of interest such as the local density, the velocity field variations, the dynamic pressure and the energy. Recall that this information is not easily accessible by means of measurements.

The present paper is organized as follows. In Section \ref{sec:model} we present the governing equations and some constitutive relations. Special attention is payed to the presence of strong density gradients in the flow and to the kinetic energy balance of the resulting system. The next Section \ref{sec:num} contains a brief description of the numerical methods and numerous computation results are presented. Finally, this paper is ended by outlining main conclusions and a few perspectives for future studies (see Section \ref{sec:concl}).

\section{Mathematical model}\label{sec:model}

In the present study we assume that an avalanche is a two-fluid flow formed by air and snow particles in suspension. The whole system moves under the force of gravity. For simplicity we assume that the mixture is a Newtonian fluid. The last assumption is not so restrictive as it can appear. The flow under consideration is such that the Reynolds number is very high ($\Re \sim 10^9$). Therefore, the transient behaviour is essentially governed by the convective terms and not by the fluid rheology. On the contrary, the rheology is very important in the flowing regime.

%%% Volume fraction %%%

% Volume dV figure %

\begin{figure}
	\centering
	\psfrag{O1}{$d\Omega_1$}
	\psfrag{O2}{$d\Omega_2$}
		\includegraphics[width=0.3\textwidth]{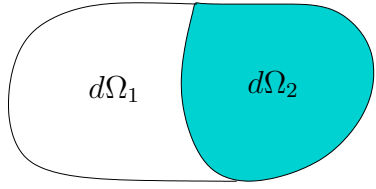}
	\caption{An elementary fluid volume $d\Omega$ occupied by two phases.}
	\label{fig:volumefract}
\end{figure}

In two-fluid flows it is natural to operate with the so-called volume fractions \cite{Ishii1975, Toumi1996, Toumi1999, Ghidaglia2001}. Consider an elementary fluid volume $d\Omega$ surrounding an interior point $P\in d\Omega$. Let us assume that the first fluid occupies volume $d\Omega_1\subseteq d\Omega$ and the second the volume $d\Omega_2\subseteq d\Omega$ (see Figure \ref{fig:volumefract}) such that
\begin{equation}\label{eq:dOmega}
  |d\Omega| \equiv |d\Omega_1| + |d\Omega_2|.
\end{equation}
The volume fraction of the fluid $i=1,2$ in the point $P$ is defined as
\begin{equation*}
  \phi_i(P) := \lim\limits_{\stackrel{|d\Omega|\to 0}{P \in d\Omega}}
  \frac{|d\Omega_i|}{|d\Omega|}.
\end{equation*}
From relation (\ref{eq:dOmega}) it is obvious that $\phi_1(P) + \phi_2(P) \equiv 1$, for any point $P$ in the fluid domain. Henceforth, it is sufficient to retain only the heavy fluid volume fraction $\phi_1$, for example, which will be denoted just by $\phi$, for the sake of simplicity.

If we assume that constant densities and kinematic viscosities of the heavy and light fluids are respectively $\rho^\pm$ and $\nu^\pm$, the mixture density $\rho$ and the dynamic viscosity $\mu$ are defined as follows:
\begin{equation}\label{eq:density}
	\rho = \phi\rho^+ + (1-\phi)\rho^-, \quad
	\mu = \phi\rho^+\nu^+ + (1-\phi)\rho^-\nu^-.
\end{equation}
Hereafter we assume that $\rho^+$ is different from $\rho^-$.

After some simple algebraic computations, the mixture dynamic viscosity $\mu$ can be expressed in terms of the mixture density $\rho$ as:
\begin{equation}\label{eq:mixrheol}
  \mu = \mu_0 + \nub\rho,
\end{equation}
where $\mu_0$ has the dimension of the dynamic viscosity $[kg/(m\cdot s)]$ and $\nub$ scales with the kinematic one $[m^2/s]$. These coefficients are related to the densities and kinematic viscosities of constitutive fluids in this way:
\begin{equation*}
  \mu_0 := \frac{\nu^-\rho^-\rho^+ - \nu^+\rho^+\rho^-}{\rho^+ - \rho^-}, \quad
  \nub := \frac{\nu^+\rho^+ - \nu^-\rho^-}{\rho^+ - \rho^-}.
\end{equation*}
Compressible Navier-Stokes equations with viscosities of the form \eqref{eq:mixrheol} were studied mathematically in \cite{Sy2005}.

\begin{rem}
In view of our applications, the snow kinematic viscosity $\nu^+$ can be parameterized as a function of temperature $T$ and snow density $\rho^+$ according to \cite{Domine2007}:
\begin{equation*}
   \nu^+ = \frac{\mu_0}{\rho^+} e^{-\alpha T} e^{\beta\rho^+},
\end{equation*}
where $\mu_0 = 3.6\times 10^6\; N\cdot s\cdot m^{-2}$, $\alpha = 0.08\; K^{-1}$, $\beta = 0.021\; m^3/kg$.
\end{rem}

Now we can state the governing equations of our physical problem. In this study we assume that both phases are constrained to have the same velocity variable $\u$. This assumption is not very restrictive. The formal justification of single-velocity two-phase models can be found in \cite{Murrone2005, Meyapin2009a, Meyapin2009}. Also, this type of models has already been successfully applied to a number of practical problems \cite{Dutykh2007a, Dias2008a, Bredmose2009, Dias2008, Dias2008b}.

The flow is assumed to be isentropic. The mass and momentum conservation equations have the classical form:
\begin{equation}\label{eq:massu}
  \dt\rho + \div(\rho\u) = 0,
\end{equation}
\begin{equation}\label{eq:momentumu}
  \dt(\rho\u) + \div(\rho\u\otimes\u) + \grad p = \div\bigl(2\mu\D(\u)\bigr) + \rho\g,
\end{equation}
where $\g$ is the acceleration due to gravity, $p$ is the hydrodynamic pressure and $\D(\u) = \frac12\bigl(\grad\u + (\grad\u)^t\bigr)$ is the strain rate tensor. 

The fluid mixing is taken into account by Fick's type law \cite{Fick1855, Fick1855a} resulting in the following quasi-compressible equation:
\begin{equation}\label{eq:divu}
  \div\u = -\div(\kappa\grad\log\rho),
\end{equation}
where the coefficient $\kappa$ has the dimension of kinematic viscosity and will be defined below. For the moment, we assume $\kappa$ to be constant, consequently, the right-hand side of equation \eqref{eq:divu} can be equivalently rewritten as $-\kappa\Delta\log\rho$.

\begin{rem}
Examples of closures similar to \eqref{eq:divu} of the form 
\begin{equation*}
  \div\u = \pm\div(\grad\phi(\rho))
\end{equation*} 
may be found in \cite{Graffi1955, Kazhikov1977, BeiraodaVeiga1983, Majda1984, Antonsev1990, Franchi2001} (the sign $\pm$ depends on monotonicity properties of the function $\phi(\rho)$). It may lead to models of low Mach number combustion \cite{Majda1984} with $\phi(\rho) = \displaystyle{\frac{1}{\rho}}$, salt or pollutant motion in a shallow layer \cite{Franchi2001} and other practical situations involving highly inhomogeneous fluids \cite{BeiraodaVeiga1982, BeiraodaVeiga1983, Bresch2007c}.
\end{rem}

\begin{rem}
For the sake of convenience we choose a cartesian coordinate system with horizontal axes directed down the slope, while the vertical axis points upwards so that to form a right-handed coordinate system. In this setting the gravity acceleration vector takes the following form:
\begin{equation*}
  \g = (g\sin\theta, -g\cos\theta),
\end{equation*}
where $\theta$ is the slope angle and $g := |\g| = 9.81$ $m/s^2$ is the standard gravity acceleration constant.
\end{rem}

\subsection{Model based on fluid volume velocity}

In this section we transform governing equations \eqref{eq:massu}, \eqref{eq:momentumu} and \eqref{eq:divu} to operate with more representative physical variables giving also a more convenient mathematical form.

Namely, we introduce the new velocity variable defined as
\begin{equation}\label{eq:uvchange}
  \v := \u + \kappa\grad\log\rho,
\end{equation}
which is sometimes referred in the literature as \textit{the mean volume velocity} (cf. \cite{Franchi2001}) or \textit{the fluid volume velocity} (cf. \cite{Brenner2005}). In this study we will retain the last term. It has been pointed out recently \cite{Brenner2005a, Brenner2006} that the fluid volume velocity $\v$ is more pertinent for flows involving high density gradients. Now, let us rewrite the system \eqref{eq:massu}, \eqref{eq:momentumu} and \eqref{eq:divu} in terms of new variable $\v$.

First of all, from Fick's law \eqref{eq:divu} it follows immediately that the flow is incompressible within the fluid volume velocity $\v$:
\begin{equation*}
  \div\v = 0.
\end{equation*}

The mass conservation equation \eqref{eq:massu} takes the following simple form:
\begin{equation*}
  \dt\rho + \div(\rho\v) = \div(\kappa\rho\grad\log\rho).
\end{equation*}
By taking into account the fact that $\kappa$ is constant and the field $\v$ is divergence-free ($\div(\rho\v) = \v\cdot\grad\rho$), we can rewrite the last equation as:
\begin{equation}\label{eq:massv}
  \dt\rho + \v\cdot\grad\rho = \kappa\Delta\rho.
\end{equation}
The latter equation is of parabolic type. The diffusive term comes from the Fick's law governing the mixing process between two fluids. In the case when $\kappa = 0$, the initial sharp interface between two phases would be simply advected by the velocity field $\v$ (coinciding with $\u$, when $\kappa \equiv 0$), thus, preventing any mixing. Henceforth, we consider the case $\kappa > 0$.

\begin{rem}
The mass conservation equation \eqref{eq:massv} can be equivalently rewritten in terms of the volume fraction $\phi$ using the mixture density representation \eqref{eq:density}:
\begin{equation}\label{eq:fickphi}
  \dt\phi + \v\cdot\grad\phi = \kappa\Delta\phi.
\end{equation}
\end{rem}

Finally, we have to transform the momentum conservation equation \eqref{eq:momentumu} according to the change of the velocity variable \eqref{eq:uvchange}. This operation will require several computations briefly presented below. For convenience, we work with equation \eqref{eq:momentumu} recasted in non-conservative form using the mass conservation \eqref{eq:massu}:
\begin{equation}\label{eq:momentu_ncv}
  \rho\dt\u + \rho(\u\cdot\grad)\u + \grad p = \rho\g + \div(2\mu\D(\u)).
\end{equation}
The three terms $\rho\dt\u$, $\rho(\u\cdot\grad)\u$ and $\div(2\mu\D(\u))$ involving $\u$ have to be rewritten:
\begin{eqnarray}\label{eq:rtu}
  \rho\dt\u &\equiv & \rho\dt\v - \kappa\rho\dt\bigl(\frac{\grad\rho}{\rho}\bigr), \\
  \rho(\u\cdot\grad)\u &\equiv & \rho(\v\cdot\grad)\v - \kappa\rho(\grad\log\rho\cdot\grad)\v - \kappa\rho(\v\cdot\grad)\grad\log\rho \\
  && + \kappa^2\rho(\grad\log\rho\cdot\grad)\grad\log\rho, \\
  \div(2\mu\D(\u)) &\equiv & \div(2\mu\D(\v)) 
  - \kappa\div(2\mu\grad\grad\log\rho).
\end{eqnarray}
In order to obtain the evolution equation for the quantity $\displaystyle{\frac{\grad\rho}{\rho}}$ arising in \eqref{eq:rtu}, we use the mass conservation equation \eqref{eq:massv}:
\begin{equation*}
  \rho\dt\bigl(\frac{\grad\rho}{\rho}\bigr) +
  \grad\bigl(\v\cdot\grad\rho\bigr) - (\v\cdot\grad\rho)\frac{\grad\rho}{\rho} 
  = \kappa\Delta\grad\rho - \kappa\Delta\rho\frac{\grad\rho}{\rho}.
\end{equation*}
Consequently, relation \eqref{eq:rtu} can be rewritten using the last result:
\begin{equation*}
  \rho\dt\u \equiv \rho\dt\v + \kappa\grad(\v\cdot\grad\rho) 
  - \kappa(\v\cdot\grad\rho)\frac{\grad\rho}{\rho} + \kappa^2\frac{\grad\rho}{\rho}\Delta\rho - \kappa^2\Delta\grad\rho.
\end{equation*}

After all these developments, the momentum conservation equation \eqref{eq:momentu_ncv} becomes:
\begin{multline*}
  \rho\dt\v + \rho(\v\cdot\grad)\v + \grad p + \underbrace{\kappa\grad(\v\cdot\grad\rho) 
  - \kappa(\v\cdot\grad\rho)\frac{\grad\rho}{\rho}}_{(I)} + \underbrace{\kappa^2\frac{\grad\rho}{\rho}\Delta\rho}_{(II)} 
  - \kappa^2\Delta\grad\rho \\
  - \kappa\rho(\grad\log\rho\cdot\grad)\v - \underbrace{\kappa\rho(\v\cdot\grad)\grad\log\rho}_{(I)}  + \underbrace{\kappa^2\rho(\grad\log\rho\cdot\grad)\grad\log\rho}_{(II)} = \\ \rho\g + \div(2\mu\D(\v)) - \kappa\div(2\mu\grad\grad\log\rho).
\end{multline*}
One can remark that terms in groups (I) and (II) can be simplified to give $\kappa{}^{t}\grad\v\,\grad\rho$ and $\kappa^2\div\bigl(\frac{\grad\rho}{\rho}\otimes\rho\bigr)$ correspondingly:
\begin{multline}\label{eq:valmost}
  \rho\dt\v + \rho(\v\cdot\grad)\v + \grad p + \kappa{}^{t}\grad\v\,\grad\rho 
  - \kappa\grad\rho\grad\v
  - \underbrace{\kappa^2\Delta\grad\rho + 
  \kappa^2\div\bigl(\frac{\grad\rho}{\rho}\otimes\rho\bigr)}_{(*)} = \\
  \rho\g + \div(2\mu\D(\v)) 
  - \underbrace{2\nub\kappa\div(\rho\grad\grad\log\rho)}_{(*)} 
  - \underbrace{2\kappa\mu_0\div(\grad\grad\log\rho)}_{(**)}.
\end{multline}
In order to obtain last two terms on the right-hand side of \eqref{eq:valmost}, expression \eqref{eq:mixrheol} for the mixture dynamic viscosity $\mu$ was used.

However, equation \eqref{eq:valmost} can be further simplified if we make a special choice for the constant $\kappa$ arising in the Fick's law \eqref{eq:divu}. More specifically, we take $\kappa \equiv 2\nub$, where $\nub$ is the mixture kinematic viscosity defined in equation \eqref{eq:mixrheol}. For this choice of Fick's diffusion coefficient $\kappa$, the terms marked with $(*)$ in \eqref{eq:valmost} disappear, since it is straightforward to check the following differential identity:
\begin{equation*}
  \div(\rho\grad\grad\log\rho) \equiv \Delta\grad\rho 
  - \div\bigl(\frac{\grad\rho}{\rho}\otimes\rho\bigr).
\end{equation*}
Finally, the term (**) in equation \eqref{eq:valmost} can be written as a gradient:
\begin{equation*}
  \div(\grad\grad\log\rho) \equiv \grad\Delta\log\rho.
\end{equation*}
Consequently, it can be incorporated in the definition of the pressure:
\begin{equation}\label{eq:pidef}
  \pi (\x,t) := p(\x, t) + 4\nub\mu_0\Delta\log\rho.
\end{equation}

If we summarize developments made above, we can state the governing equations of the proposed model:
\begin{eqnarray}\label{eq:incomprv}
  \div\v &=& 0, \\
  \dt\rho + \v\cdot\grad\rho &=& 2\nub\Delta\rho, \label{eq:massv2} \\
  \rho\dt\v + \rho(\v\cdot\grad)\v + \grad\pi + 2\nub{}^{t}\grad\v\,\grad\rho 
  - 2\nub\grad\rho\grad\v &=&
  \rho\g + \div(2\mu\D(\v)) \label{eq:momentumv}
\end{eqnarray}
Equations \eqref{eq:incomprv}, \eqref{eq:massv2} and \eqref{eq:momentumv} have to be completed by appropriate initial and boundary conditions to form a well-posed problem.

Initially, the velocity, pressure and density (or equivalently, volume fraction) fields have to be imposed. Concerning boundary conditions, the usual no-slip condition $\v=\vec{0}$ can be imposed. Another possibility is to impose the following partial slip condition:
\begin{equation}\label{eq:partialB}
  \v\cdot\n=0, \quad \bigl((1-\alpha)\v + \alpha(\D(\v)\cdot\n)\bigr)\cdot\ttau = 0,
\end{equation}
where $\ttau$ is the tangent vector to the boundary and $\alpha\in[0,1]$ is the friction parameter. We have to say that the latter is known to be physically more relevant \cite{Etienne2004a}.

If it is required by a computational algorithm as in our case, for example, governing equations can be recast in the conservative form. Using the incompressibility condition \eqref{eq:incomprv}, the mass conservation equation \eqref{eq:massv2} becomes:
\begin{equation*}
  \dt\rho + \div(\rho\v) = \div(2\nub\grad\rho).
\end{equation*}
If we sum up equation \eqref{eq:momentumv} with mass conservation \eqref{eq:massv2} multiplied by $\v$, we will obtain the following equation with advective terms written in the conservative form:
\begin{multline}\label{eq:momentcons}
  \dt(\rho\v) + \div(\rho\v\otimes\v) + \grad\pi + 2\nub{}^{t}\grad\v\,\grad\rho 
  - 2\nub\grad\rho\grad\v = \\
  \rho\g + 2\nub\v\,\Delta\rho + \div(2\mu\D(\v)).
\end{multline}
The dimensionless form and discretization of these equations will be discussed briefly below in Sections \ref{sec:dimension} and \ref{sec:num} correspondingly.

\subsection{Kinetic energy evolution}

In this section we would like to derive an integral identity which describes the kinetic energy evolution associated to the system \eqref{eq:incomprv} -- \eqref{eq:momentumv}. Throughout all developments in this section we assume the no-slip condition $\v = 0$ on the fluid domain $\Omega$ boundary $\partial\Omega$. The same result will hold if we assume periodic boundaries or appropriate decay conditions at infinity.

First, we multiply the mass conservation equation \eqref{eq:massv2} by $\displaystyle{\frac{|\v|^2}{2}}$, momentum conservation \eqref{eq:momentumv} by $\v$, sum them up and integrate over the fluid domain $\Omega$. After a few integrations by part, using the incompressibility \eqref{eq:incomprv} and boundary conditions, we obtain the following identity:
\begin{multline*}
  \dt\int\limits_{\Omega}\rho\frac{|\v|^2}{2}\,d\x =
   \int\limits_{\Omega}\rho\g\cdot\v\,d\x +
   \underbrace{\int\limits_{\Omega}2\nub\Delta\rho\frac{|\v|^2}{2}\,d\x +
   \int\limits_{\Omega}2\nub(\grad\rho\grad\v)\v\,d\x}_{(I)} + \\
   \underbrace{\int\limits_{\Omega}\div(2\nub\rho\D(\v))\v\,d\x - 
   \int\limits_{\Omega}2\nub({}^t\grad\v\grad\rho)\v\,d\x}_{(II)} + 
   \underbrace{\int\limits_{\Omega}\div(2\mu_0\D(\v))\v\,d\x}_{(III)}
\end{multline*}
Two terms from the group (I) cancel each other after one integration by parts:
\begin{equation*}
	\int\limits_{\Omega}2\nub\Delta\rho\frac{|\v|^2}{2}\,d\x = 
	-\int\limits_{\Omega}2\nub\grad\rho\cdot\grad\bigl(\frac{|\v|^2}{2}
	\bigr)\,d\x = 
	-\int\limits_{\Omega}2\nub(\grad\rho\grad\v)\v\,d\x.
\end{equation*}
The group (II) terms can be also transformed to give:
\begin{equation*}
  -\int\limits_{\Omega}2\nub\rho|\D(\v)|^2\,d\x + 
  \int\limits_{\Omega}2\nub\rho{}^{t}\grad\v:\grad\v\,d\x \equiv 
  -\int\limits_{\Omega}2\nub\rho|\A(\v)|^2\,d\x,
\end{equation*}
where $\A(\v) := \frac12\bigl(\grad\v - {}^{t}\grad\v\bigr)$ is the antisymmetric part of the velocity gradient $\grad\v$ and $A:B = a_{ij}b_{ij}$ is the usual contracted product of two square matrices $A = (a_{ij})_{1\leq i,j\leq n}$, $B = (b_{ij})_{1\leq i,j\leq n}$.

Finally, the term (III) similarly can be transformed as:
\begin{equation*}
  \int\limits_{\Omega}\div(2\mu_0\D(\v))\v\,d\x \equiv
  - \int\limits_{\Omega} 2\mu_0|\D(\v)|^2\,d\x
\end{equation*}

Thus, we come to the following integral identity governing the kinetic energy evolution associated to the system \eqref{eq:incomprv} -- \eqref{eq:momentumv} (in closed conditions):
\begin{equation}\label{eq:kinenergy}
  \dt\int\limits_{\Omega}\rho\frac{|\v|^2}{2}\,d\x =
  \int\limits_{\Omega}\rho\g\cdot\v\,d\x
  - \underbrace{\int\limits_{\Omega} 2\mu_0|\D(\v)|^2\,d\x}_{(a)}
  - \underbrace{\int\limits_{\Omega}2\nub\rho|\A(\v)|^2\,d\x}_{(b)}.
\end{equation}
In this integral equality each term has a specific physical sense. The term on the left-hand side represents the rate of kinetic energy change per a unit of time. On the right-hand side the first term is the work done by the gravity force, thus, describing the energy input into the system. Two last terms represent various dissipative processes. The first term (a) can be seemingly ascribed to the dissipation due to viscous forces, while the second one (b) comes from mixing processes between two phases.

Despite the fact that governing equations \eqref{eq:incomprv} -- \eqref{eq:momentumv} are more complicated than classical Navier-Stokes equations, due to the judicious choice of Fick's constant $\kappa$ we were able to get a kinetic energy balance in simple and physically sound form \eqref{eq:kinenergy}.

\subsection{Dimensional analysis}\label{sec:dimension}

In this section we perform a dimensional analysis of the governing equations \eqref{eq:incomprv} -- \eqref{eq:momentumv} in order to reveal important scaling parameters. Henceforth, starred variables denote dimensional quantities throughout this section.

The initial avalanche height $h_0$ and the heavy fluid density $\rho^+$ are chosen to be the characteristic length and density scales correspondingly. The velocity field is adimensionalized by a typical flow speed $u_0$. Finally, from characteristic length and velocity, it is straightforward to deduce the time scale. Consequently, the scaling for the independent variables is
\begin{equation*}
  \x^* =  h_0 \x, \quad t^* = \frac{h_0}{u_0} t,
\end{equation*}
and dimensionless dependent variables $\rho, \u, p$ are introduced in this way:
\begin{equation*}
  \rho^* = \rho^+\rho, \quad \u^* = u_0\;\u, \quad
  \pi^* = \rho^+u_0^2\; \pi, \quad \mu^* = \rho^+\nub\mu
\end{equation*}

The governing equations \eqref{eq:incomprv} -- \eqref{eq:momentumv} in dimensionless form become:
\begin{eqnarray*}
  \div\v &=& 0, \\
  \dt\rho + \v\cdot\grad\rho &=& \frac{1}{\Re}\Delta\rho, \\
  \rho\dt\v + \rho(\v\cdot\grad)\v + \grad\pi 
  + \frac{1}{\Re}{}^{t}\grad\v\,\grad\rho 
  - \frac{1}{\Re}\grad\rho\grad\v &=&
  \frac{1}{\Fr^2}\rho\g + \frac{1}{\Re}\div(\mu\D(\v))
\end{eqnarray*}
This procedure reveals two important scaling parameters -- the Reynolds number $\Re$ \cite{Reynolds1883} and the Froude number $\Fr$ \cite{Belanger1828} which are defined as
\begin{equation*}
  \Re := \frac{h_0u_0}{2\nub}, \qquad
  \Fr := \frac{u_0}{\sqrt{gh_0}}.
\end{equation*}
Recall that the Reynolds number $\Re$ gives a measure of the ratio of inertial forces to viscous forces. In practice this number characterizes different flow regimes, such as laminar or turbulent flow. The Froude number is a ratio of inertia and gravitational forces. It is a hydrodynamic equivalent of the Mach number. For free-surface flows it specifies the nature of the flow (subcritical or supercritical) \cite{Dias2002}.

\begin{rem}
We would like to discuss the difference between the physical pressure field $p(\x,t)$ and the modified pressure $\pi(\x,t)$. If we turn to dimensionless variables in equation \eqref{eq:pidef}, we will obtain the following relation:
\begin{equation*}
  \pi(\x,t) = p(\x,t) + \frac{\mu}{\Re^2}\Delta\log\rho.
\end{equation*}
Taking into account the typical values of the Reynolds number in our applications, we can conclude that this difference is completely negligible, at least from the practical point of view.
\end{rem}

In fact, there are two additional scaling parameters $\delta$ and $\lambda$ hidden in definitions of the density $\rho$ and dynamic viscosity of the mixture:
\begin{equation*}
  \rho := \frac{\rho^*}{\rho^+} = \phi + (1-\phi)\delta, \qquad \delta := \frac{\rho^-}{\rho^+},
\end{equation*}
\begin{equation*}
  \mu := \frac{\mu^*}{\rho^+\nub} = 
  (1-\delta)\phi + \frac{\delta\lambda(1-\delta)}{1-\delta\lambda}, \qquad 
  \lambda := \frac{\nu^-}{\nu^+},
\end{equation*}
where we substituted the following representation of the Fick's coefficient:
\begin{equation}\label{eq:nub}
  \nub = \nu^+\frac{1-\delta\lambda}{1-\delta}.
\end{equation}

Actually, the densities ratio $\delta$ can be related to the well-known Atwood number $\At$ (cf. \cite{Glimm2001, Likhachev2005}):
\begin{equation*}
  \At := \frac{\rho^+ - \rho^-}{\rho^+ + \rho^-} = \frac{1-\delta}{1+\delta}\;.
\end{equation*}
The powder-snow avalanche regime is characterized by very high values of Reynolds number $\Re$ and low density ratios $\delta$ \cite{McClung1993, Ancey2003}:
\begin{equation*}
  \Re \sim 10^6, \qquad 0.05 \leq \delta \leq 0.25.
\end{equation*}
Thus, the flow is clearly turbulent. Nevertheless, in the present study we do not consider any turbulence modeling beyond the scales resolved by the numerical method.

There is much less available information on the snow viscosity and snow rheology, in general \cite{Domine2007}. The viscosity ratio parameter $\lambda$ can be related (using equation \eqref{eq:nub}) to the so-called Schmidt number $\Sc$ which represents the ratio of the fluid viscosity to mass diffusivity:
\begin{equation*}
  \Sc := \frac{\nu^+}{2\nub} = \frac{1-\delta}{2(1-\delta\lambda)}
\end{equation*}
This number was named after the German engineer Ernst Heinrich Wilhelm Schmidt (1892 -- 1975) and it is used to characterize fluid flows with simultaneous momentum and mass diffusion processes \cite{Yeung2002, Schumacher2003}. For powder-snow avalanches, M. Cl\'{e}ment-Rastello reported \cite{Clement-Rastello2001} the following values of the Schmidt number:
\begin{equation*}
  0.5 \leq \Sc \leq 1.
\end{equation*}

When the Reynolds number is sufficiently high (inertial regime), two scaling parameters to be respected are the density ratio $\delta = \frac{\rho^-}{\rho^+}$ and the Froude number $\Fr := \frac{u_0}{\sqrt{gh_0}}$. Recall that in the Boussinesq regime ($\delta\to 0$), only the Froude number has to be respected. We quote \cite{Primus2004} reporting on this important issue:
\begin{quote}
  \dots Satisfying the Froude number and density ratio similarities in the laboratory means that a very high velocity is necessary, which calls for a very large channel. It is not possible to satisfy the density ratio similarity, because dimensionless number differs by several orders of magnitude between the processes that unfold in nature and those reproduced in the laboratory\dots
\end{quote}
As a result, the interpretation of laboratory results is quite ambiguous. At this point, numerical simulation should be considered as a complementary tool to physical modeling.

\section{Numerical methods and simulation results}\label{sec:num}

In this article we perform Direct Numerical Simulations (DNS) of a snow cloud moving down a steep slope. Our solver is based on a freely available CFD toolbox \OpenFOAM. All computations performed in this study are 3D with only one cell in $z$ direction for the sake of efficiency. In principle, the extension to truly 3D configurations is possible with this code.

In order to implement model \eqref{eq:incomprv} -- \eqref{eq:momentumv}  we modified the standard solver \textsf{twoLiquidMixingFoam} which discretizes incompressible two-fluid Navier-Stokes equations with Fick's diffusion term in the volume fraction transport equation \eqref{eq:fickphi}. The main modification concerns the momentum balance equation. Namely, we had to incorporate two nonconservative terms $2\nub{}^{t}\grad\v\,\grad\rho$ and $2\nub\grad\rho\grad\v$. For information, we provide here a piece of the code written in the internal \OpenFOAM language which corresponds to the discretization of equation \eqref{eq:momentcons}:
\begin{verbatim}
fvVectorMatrix UEqn
(
  fvm::ddt(rho, U) + 
  fvm::div(rhoPhi, U) -
  fvm::laplacian(muf, U) -
  U*fvc::div(DabRho*(mesh.Sf()&fvc::interpolate(fvc::grad(rho))))+
  DabRho*(fvc::grad(U)().T() & fvc::grad(rho)) - 
  DabRho*(fvc::grad(U)() & fvc::grad(rho)) - 
  fvc::div(muf*(mesh.Sf() & fvc::interpolate(fvc::grad(U)().T())))
);
\end{verbatim}

Time derivatives were discretized with the classical implicit Euler scheme. An upwind second order finite volume method was employed in space. For more details on the retained discretization scheme we refer to \cite{Jasak1996, Rusche2002, OpenFOAM2007}. The choice of the finite volume method is justified by its excellent stability and local conservation properties (especially in comparison to FEM \cite{Etienne2004, Etienne2004a, Etienne2005}).

\subsection{Description of numerical computations}

\begin{figure}
  \centering
  \psfrag{H0}{$h_0$}
  \psfrag{L}{$\ell$}
  \psfrag{g}{$\g$}
  \psfrag{H}{h}
  \psfrag{Hs}{$h_s$}
  \psfrag{X}{$x$}
  \psfrag{Y}{$y$}
  \psfrag{Z}{$z$}
  \includegraphics[width=0.99\textwidth]{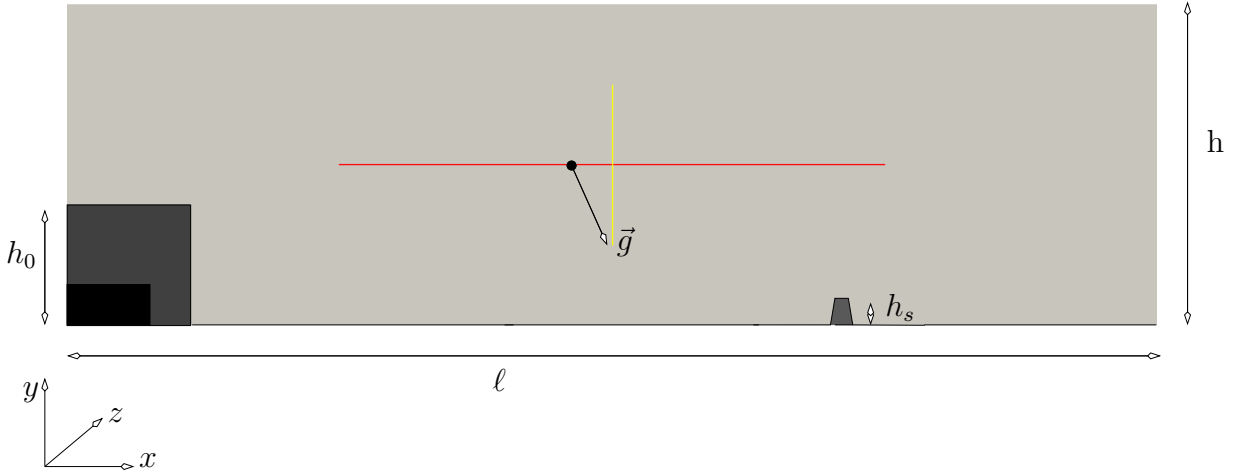}
  \caption{Sketch of the computational domain and initial condition description.}
  \label{fig:setup}
\end{figure}

In this study we try to reproduce \textit{in silico} a classical lock-exhange type flow with an obstacle placed in the computational domain. The sketch of the fluid domain and the initial condition description are given on Figure \ref{fig:setup}. Experiment consists in releasing a heavy fluid which flows under the gravity force along an inclined channel. The values of all physical parameters are provided in Table \ref{tab:params}, while computational domain and discretization parameter are given in Table \ref{tab:paramsDomain}.

\begin{table}
     \begin{center}
       \begin{tabular}{c|l}
         \hline\hline
         
         \hline
         \textit{parameter} & \textit{value} \\
         \hline
         gravity acceleration $g$, $m/s^2$ & $9.8$ \\ 
         \hline
         slope, $\theta$ & $32^\circ$ \\
         \hline
         friction parameter, $\alpha$ & 0.3 \\
         \hline
         heavy fluid density, $\rho^+$, $kg/m^3$ & 20 \\
         \hline
         light fluid density, $\rho^-$, $kg/m^3$ & 1 \\
         \hline
         heavy fluid kinematic viscosity, $\nu^+$, $m/s^2$ & 4.8$\times 10^{-4}$\\
         \hline
         light fluid kinematic viscosity, $\nu^-$, $m/s^2$ & 1.0$\times 10^{-4}$\\
         \hline\hline
       \end{tabular}
       \caption{Values of various parameters used for numerical simulations. The friction parameter is used in the partial slip boundary condition \eqref{eq:partialB}.}
       \label{tab:params}
     \end{center}
\end{table}

\begin{table}
    \begin{center}
       \begin{tabular}{c|l}
         \hline\hline
         
         \hline
         \textit{parameter} & \textit{value} \\
         \hline
         domain height, $h$, m & 0.8 \\
         \hline
         total length, $\ell$, m & 2.7 \\
         \hline
         mesh height, $\Delta y$, m & 0.0026 \\
         \hline
         mesh length, $\Delta x$, m & 0.0033 \\
         \hline
         initial avalanche height, $h_0$, m & 0.3 \\
         \hline
         obstacle height, $h_s$, m & $0.2 h_0$ \\
         \hline
         obstacle thickness, m & 0.04 \\
         \hline
         
         \hline\hline
       \end{tabular}
       \caption{Dimensions of the configuration and mesh.}
       \label{tab:paramsDomain}
     \end{center}
\end{table}

The objective of this study is twofold. Besides presenting numerical results on a gravity-driven two-fluid flow, we also would like to shed some light onto the interaction process between an avalanche and an obstacle. At the present stage we assume the obstacle to be absolutely rigid but this assumption can be relaxed in future investigations. Results presented here are obtained for the obstacle height $h_s = 0.2h_0$, where $h_0$ is the initial mass height. We would like to underline also the presence of the stratification in our initial condition. We assume that there is a dense core of the height $h_0/3$ composed of the pure heavy fluid ($\phi = 1$, consequently $\rho = \rho^+$), which is surrounded by a lighter snow layer with the volume fraction $\phi = 0.4$. All results presented below are computed for simplicity with the usual no-slip boundary condition $\v=\vec{0}$. However, we performed several tests with the partial slip condition \eqref{eq:partialB} leading generally to better results with respect to the snow entrainment at the bottom. Moreover, the friction coefficient $\alpha$ allows for some \textit{tuning} depending on local soil conditions.

\subsection{Simulation results}

%%%%%%% Volume fraction %%%%%%%%%%%

\begin{figure}
  \centering
  \includegraphics[width=0.99\textwidth]{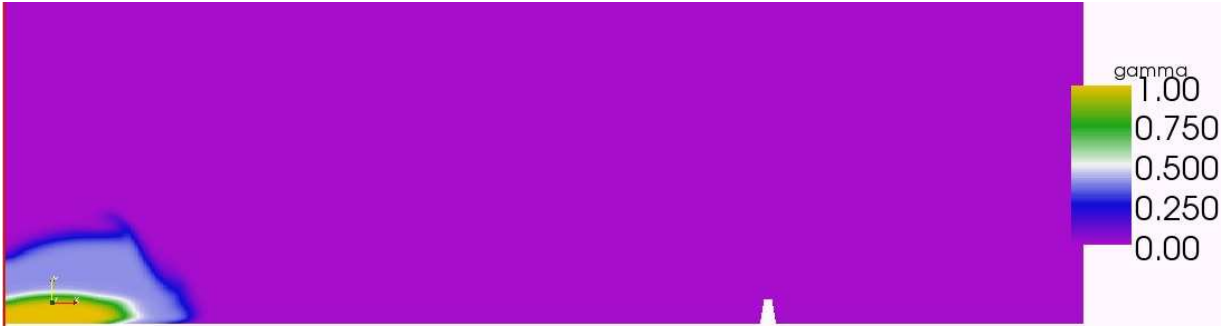}
  \caption{Avalanche at $t = 10$ $s$. The color scale ranges from $0$ to the maximum value $1.0$.}
  \label{fig:GammaT10}
\end{figure}

\begin{figure}
  \centering
  \includegraphics[width=0.99\textwidth]{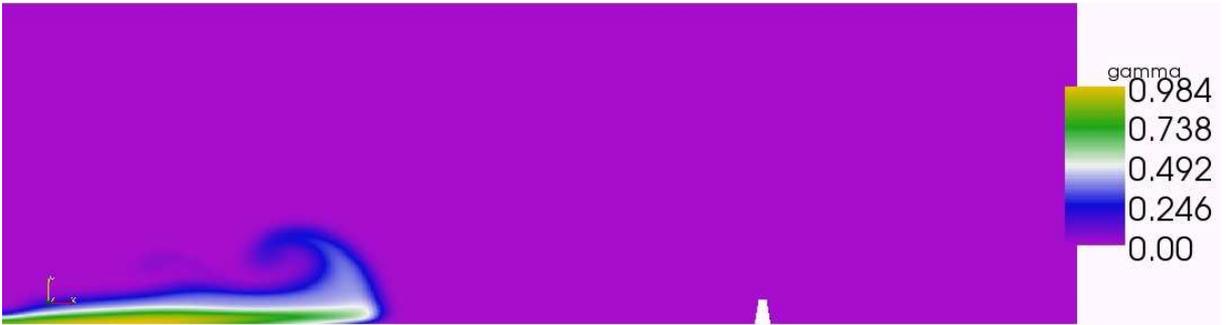}
  \caption{Avalanche at $t = 25$ $s$. The color scale ranges from $0$ to the maximum value $0.984$.}
  \label{fig:GammaT25}
\end{figure}

\begin{figure}
  \centering
  \includegraphics[width=0.99\textwidth]{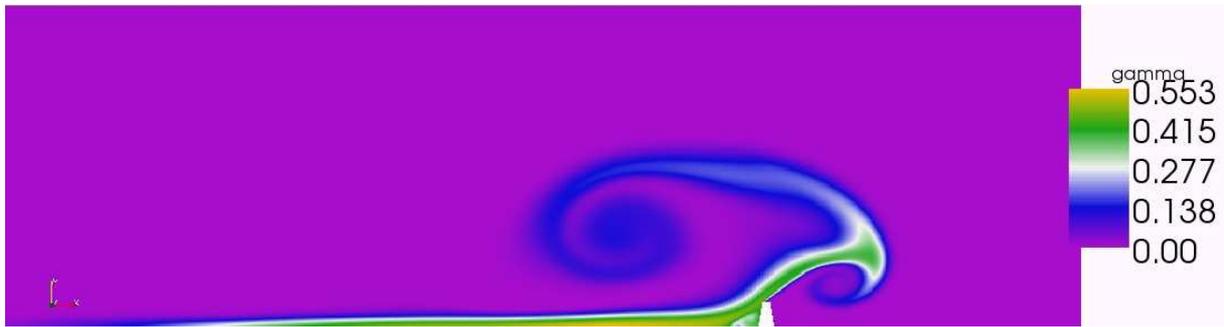}
  \caption{Avalanche at $t = 60$ $s$. The color scale ranges from $0$ to the maximum value $0.553$.}
  \label{fig:GammaT60}
\end{figure}

Snapshots of the volume fraction evolution are presented on Figures \ref{fig:GammaT10} -- \ref{fig:GammaT60}. At the beginning, the initial rectangular mass gradually transforms under the force of gravity into more classical elliptic form  (see Figure \ref{fig:GammaT10}). Then, this mass enters into the sliding regime (see Figure \ref{fig:GammaT25}) until the interaction with the obstacle (see Figure \ref{fig:GammaT60}).

Our simulations clearly show that a Kelvin-Helmholtz type instability \cite{Helmholtz1868, Kelvin1871, Chandrasekhar1981, Drazin2004} develops locally during the propagation stage. Previous simulations involving Adaptive Mesh Refinement (AMR) techniques confirmed this observation \cite{Etienne2004a, Etienne2005}. On the other hand, the interaction process with the obstacle creates a jet directed upward. This jet has a mushroom-like shape typical for Rayleigh-Taylor instability \cite{Rayleigh1883, Taylor1950, Drazin2004}.

%%%%%%%%%% Velocity %%%%%%%%%%%%%

\begin{figure}
  \centering
  \includegraphics[width=0.99\textwidth]{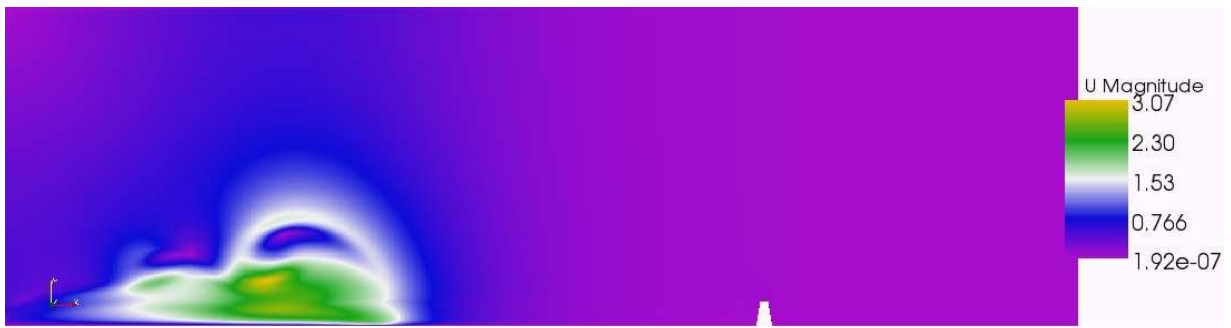}
  \caption{Velocity field magnitude at $t = 25$ $s$. The color scale ranges from $0$ $m/s$ to the maximum value $3.07$ $m/s$.}
  \label{fig:velocity25}
\end{figure}

\begin{figure}
  \centering
  \includegraphics[width=0.99\textwidth]{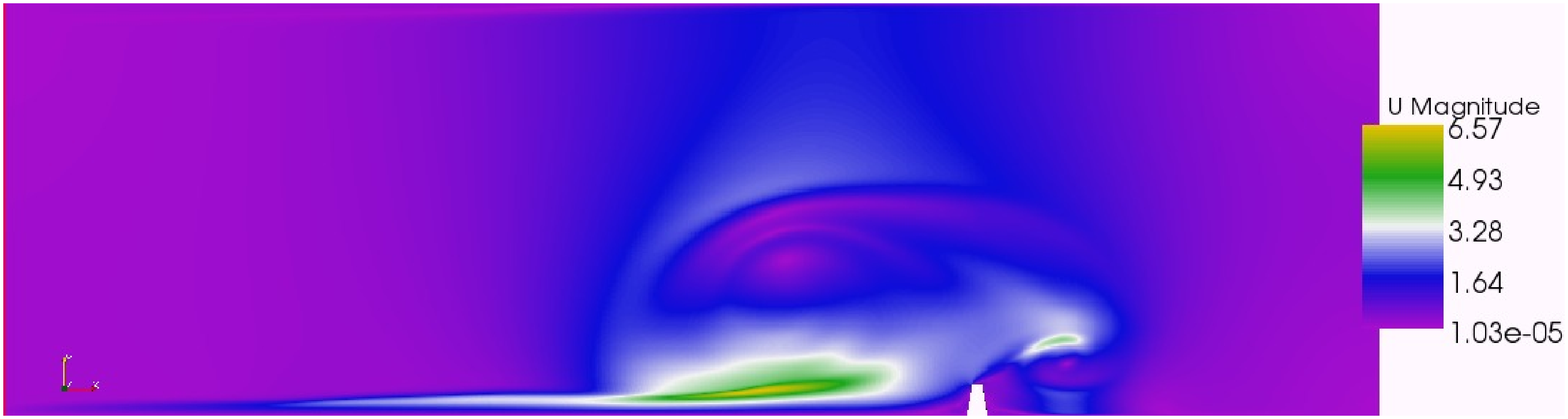}
  \caption{Velocity field magnitude at $t = 60$ $s$. The color scale ranges from $0$ $m/s$ to the maximum value $6.57$ $m/s$.}
  \label{fig:velocity60}
\end{figure}

Several authors pointed out an intriguing feature of the avalanche type flows \cite{Dufour2001, Rastello2004}. Namely, it was shown by radar measurements that the maximum velocity inside the avalanche exceeds the front velocity by 30\% -- 40\%. For this purpose we visualize the velocity field magnitude during the propagation stage (see Figures \ref{fig:velocity25} and \ref{fig:velocity60}). Qualitatively, our computations are in conformity with these experimental results.

%%%%%%%%%%% Energy %%%%%%%%%%%%%%

The same simulation was also performed with the so-called (in this study) standard model used in \cite{Etienne2004} and implemented also in \OpenFOAM as the \textsf{twoLiquidMixingFoam} solver:
\begin{eqnarray*}
  \div\v &=& 0, \\
  \dt\rho + \v\cdot\grad\rho &=& 2\nub\Delta\rho, \\
  \rho\dt\v + \rho(\v\cdot\grad)\v + \grad p &=& \rho\g + \div(2\mu\D(\v))
\end{eqnarray*}
The standard model differs from governing equations \eqref{eq:incomprv} -- \eqref{eq:momentumv} essentially by two non-conservative terms in the momentum balance equation along with small differences in the pressure definition.

\begin{figure}
  \centering
  \includegraphics[width=0.99\textwidth]{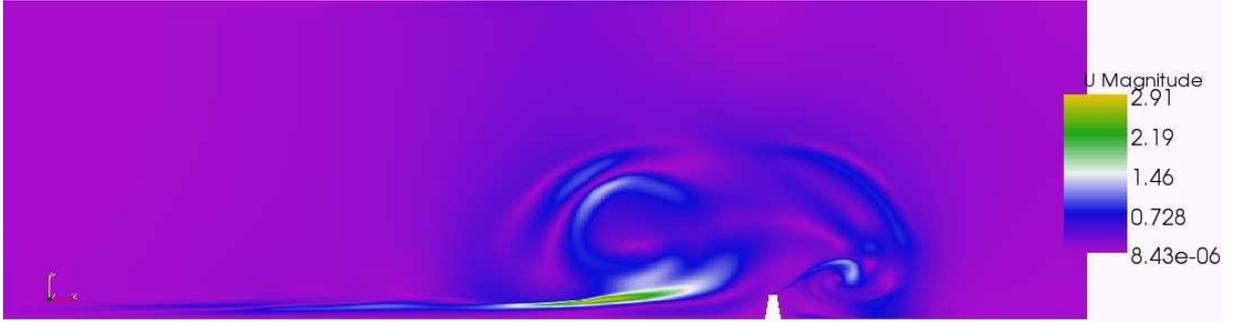}
  \caption{Magnitude of two velocity fields difference computed according to the standard and new models at $t = 60$ $s$. The color scale ranges from $0$ $m/s$ to the maximum value $2.91$ $m/s$.}
  \label{fig:diffU60}
\end{figure}

\begin{figure}
  \centering
  \includegraphics[width=0.99\textwidth]{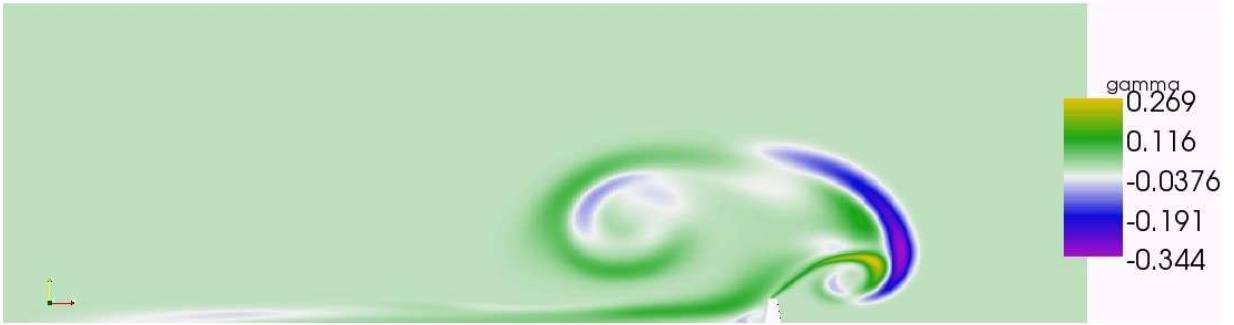}
  \caption{Difference of the volume fraction distributions computed according to the standard and new models at $t = 60$ $s$. The color scale ranges from $-0.344$ to the maximum value $0.269$.}
  \label{fig:diffGam60}
\end{figure}

The magnitude of the difference between velocity fields obtained with the standard and novel models is represented on Figure \ref{fig:diffU60}. One can see that differences are not negligible and attain their maximum value of $2.91$ $m/s$ near the lower boundary. Corresponding difference field of two volume fraction distributions are shown on Figure \ref{fig:diffGam60}. Here the maximum value of the difference $0.344$ is attained in the avalanche front.

\begin{figure}
  \centering
  \includegraphics[width=0.89\textwidth]%
  {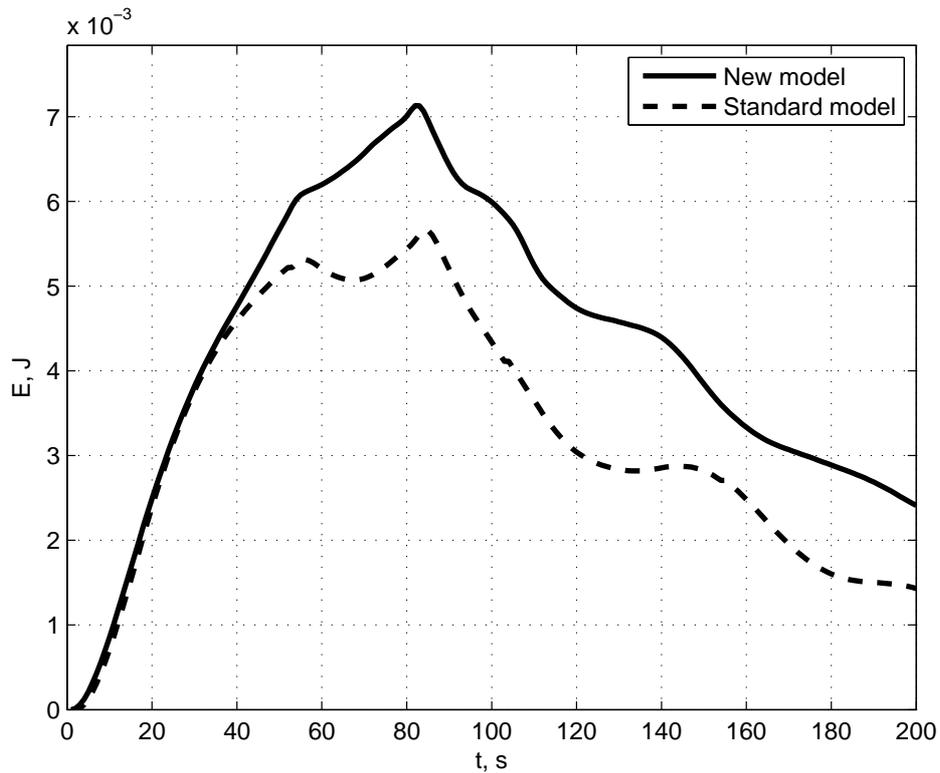}
  \caption{Kinetic energy evolution during the simulation. The continuous blue line corresponds to computations performed with equations \eqref{eq:incomprv} -- \eqref{eq:momentumv}, while the dashed black line is produced using the standard model.}
  \label{fig:kinenergy}
\end{figure}

During the simulation we also computed the kinetic energy evolution. The comparison result is presented on Figure \ref{fig:kinenergy}. The energy grows almost linearly in time during the propagation stage for both models. The differences start to appear just before the interaction process with the obstacle. These results indicate that proposed above model \eqref{eq:incomprv} -- \eqref{eq:momentumv} allows for higher energy levels under equal numerical conditions.

\begin{rem}
The idea to use the kinetic energy loss to estimate the efficiency of a dike was already proposed by Beghin and Closet in 1990 \cite{Beghin1990}. However, they had very limited information on the flow structure (especially velocity and density profiles). That is why they decided to approximate this quantity by the ratio $\abs{U^2 - U'^2}/U^2$. Here $U'$ is the front velocity at certain distance below the dike and $U$ is the front velocity of the reference avalanche measured at the same point.
\end{rem}

%%%%%% Impact pressures %%%%%%%%

\subsection{Impact pressures}

For many practical applications we have to estimate the loading exerted on a structure by an avalanche impact. Incidentally, the avalanche hazard level is attributed depending on the estimated impact pressure values \cite{Lievois2006}. Moreover, this information is crucial for the design of buildings and other structures exposed to this natural hazard.

In engineering practice, it is common to determine the impact pressures according to the following formula \cite{McClung1993}:
\begin{equation}\label{eq:simpleImpact}
  P_d = K p_{\textrm{ref}} = K \bar\rho U_f^2
\end{equation}
where $K$ is a parameter depending on the obstacle configuration, $\bar\rho$ is the average avalanche density and $U_f$ is the front velocity. For small obstacles it is advised to take $K = 1$ and for big ones $K = 2\sin\alpha$, where $\alpha$ is the incidence angle. However, as it is pointed out in \cite{Beghin1991}, it is difficult to estimate the maximum pressure exerted by an avalanche since we have only very limited information on the vertical structure of the flow.

\begin{rem}
  The given above formula (\ref{eq:simpleImpact}) is applicable to avalanches in inertial regime. When we deal with a gravity flow regime (in Fluid Mechanics we call it the Stokes flow \cite{Happel1983}), the situation is more complicated since the flow is governed by the rheology which is essentially unknown \cite{Ancey2004a}. In this case, engineers use another expression \cite{Ancey2006}:
  \begin{equation*}
    P_d = 2\bar\rho g (h-z).
  \end{equation*}
\end{rem}

For aerosol avalanches, Beghin and Closet \cite{Beghin1990} proposed the following empirical law to estimate the impact pressure:
\begin{equation*}
  P_d = \frac{K}{2}K_a(z)\bar\rho U_f^2,
\end{equation*}
where $K_a(z)$ is a dimensionless factor taking into account for the velocity variations in the upward direction. They also suggested an idealized form of the factor $K_a(z)$:
\begin{equation}\label{eq:Kaz}
  K_a(z) = \left\{ \begin{array}{cl}
  								   10, & z < 0.1h, \\
  								   19-90z, & 0.1h \leq z \leq 0.2h, \\
  								   1, & z > 0.2h,
  								 \end{array} \right.
\end{equation}
where $h$ is the impacting avalanche height. It was shown later \cite{Naaim-Bouvet2003} that this approximation underestimates the dynamic pressure in all parts of the flow.

\begin{figure}
  \centering
  \includegraphics[width=0.99\textwidth]{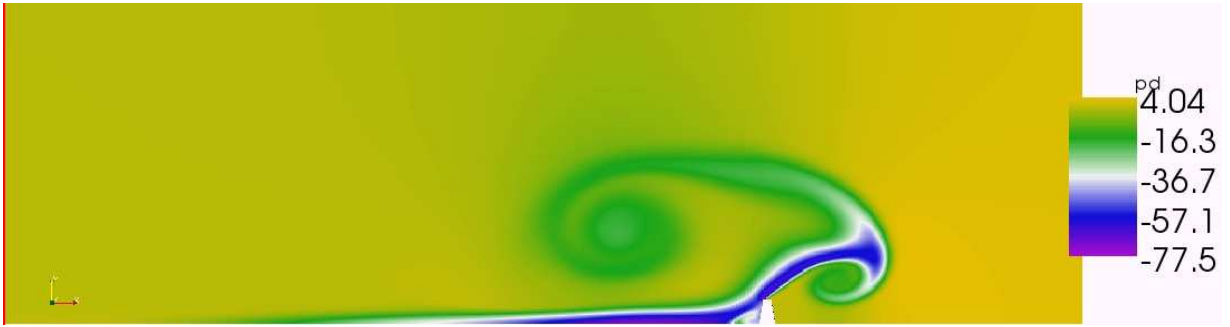}
  \caption{Dynamic pressure distribution in the computational domain at $t = 60$ $s$. The color scale ranges from $-77.5$ $Pa$ to the maximum value $4.04$ $Pa$.}
  \label{fig:PD60}
\end{figure}

The big advantage of the presented here approach is that we have the complete information on the mass but also the pressure distribution in the whole domain. On Figure \ref{fig:PD60} one can see the distribution of the dynamic pressure at $t = 60$ s during the impact process. It is important to note that an "aspiration" zone is revealed near the obstacle base along with a high pressure field at the top. In civil engineering this shear-type loading is known to be particularly dangerous for structures.

The methodology presented in this study allows to determine the impact pressures with required accuracy. The pressure profiles along the impacted wall can be easily extracted from numerical computations, thus, replacing empirical formulas such as \eqref{eq:Kaz}. Our numerical experiments show also that the form of the vertical pressure distribution does not change drastically when we vary the obstacle height in some reasonable limits. It means that precomputed pressure profiles may be scaled and reused for different obstacles within engineering accuracy \cite{Beghin1990, McClung1993}.

\section{Conclusions and perspectives}\label{sec:concl}

In the present paper we derive a novel model for the flow of two miscible inhomogeneous fluids. The mixing between two fluids is assumed to be modeled by the Fick's law \eqref{eq:divu}. Then, we reformulate the model in terms of the so-called volume-based velocity \cite{Brenner2005, Brenner2005a} in which the flow is exactly incompressible. It is also shown that the novel model allows for a physically sound evolution of the kinetic energy. 

We show some preliminary results on the numerical modeling of powder-snow avalanches. We compute the evolution of an avalanche from the beginning until hitting against a trapezoidal obstacle. Impact pressures extracted from these computations can be used for the design of protecting structures.

For the moment the obstacle is assumed to be absolutely rigid. In future studies this assumption can be relaxed and efforts exerted on the obstacle could be coupled with a solid mechanics part \cite{Keylock2001}. This research axis seems to be still underexplored. However, there is already interesting experimental material on the avalanche interaction with solid obstacles \cite{Lane-Serff1995, Primus2003, Primus2004, Naaim-Bouvet2003, Berthet-Rambaud2004}.

There is also an important question of boundary conditions. All computational results presented above were obtained with the usual no-slip condition. The development of other types of physically sound boundary conditions which lead to well-posed mathematical problems is highly needed.

Recall that a powder-snow avalanche front can reach the speed up to $100$ m/s. Consequently, the Mach number $\Ma$ attains relatively high values:
\begin{equation*}
  \Ma := \frac{u_f}{c_s} \approx 0.3,
\end{equation*}
where $c_s$ is the sound speed in the air. It means that compressible effects may be important during the propagation stage. In general, impact events are followed by strong compressions. Hence, in future works we are planning to take into account the compressibility in some weak and strong senses \cite{Dias2008a, Dias2008, Dias2008b, Dutykh2007a} and to perform the comparisons with the present results. Obviously, at this point a validation against experimental data is highly recommended.

The rheology of avalanches should be further investigated \cite{Ancey2004a} and future models will take this information into account. At this stage, close collaboration between physicists and mathematicians is needed to bring the answers on challenging questions.

\section*{Acknowledgment}

The authors would like to acknowledge the University of Savoie for the PPF grant linked to the project: ``Math\'ematiques et avalanches de neige, une rencontre possible?''. They also would like to thank Professor Carmen de JONG for interesting discussions around snow avalanches.

The support from the Research network VOR (Professors Jacky Mazars and Denis Jongmans) and Cluster Environnement through the program ``Risques gravitaires, s\'eismes'' is acknowledged.

Finally, Denys Dutykh acknowledges the support from CNRS/ASR under the project $N^\circ$ 23975 ``Solutions analytiques et num\'eriques pour les mod\`eles des avalanches de neige poudreuse''.

%%%% Bibliography  %%%%%%%%%%
\bibliographystyle{alpha}
\bibliography{biblio}

\end{document}